\documentclass{article}

\usepackage[english]{babel}

\usepackage[letterpaper,top=2cm,bottom=2cm,left=2cm,right=2cm,marginparwidth=1.75cm]{geometry}

\usepackage{mathrsfs}
\usepackage{amsmath}
\usepackage{amssymb}
\usepackage{epsfig}
\usepackage{graphicx}
\usepackage{subfigure}
\usepackage{multirow}
\usepackage{epstopdf}
\usepackage{authblk}
\usepackage{cite}

\def\be{\begin{eqnarray}}
\def\en{\end{eqnarray}}
\def\bea{\begin{aligned}}
\def\ena{\end{aligned}}
\def\prd{{Phys. Rev. D}~}

\def\non{\nonumber\\}
\title{Quasi-two-body decays $B_c\to D^*h\to D\pi h$ in the perturbative QCD}
\author{   Yan-Chao Zhao, Zhi-Qing Zhang
\footnote{Electronic address: zhangzhiqing@haut.edu.cn (corresponding author)}, Zi-Yu Zhang, Zhi-Jie Sun and Qiu-Bo Meng} 
\affil{ \it \small Institute of Theoretical Physics, School of Sciences, Henan University
of Technology,\\ \it Zhengzhou, Henan 450052, China }

\begin{document}
\maketitle

\begin{abstract}
In this work, we investigate the quasi-two-body decays $B_c\to D^*h\to D\pi h$ with $h = (K^0,\pi^0,\eta,\eta^{\prime})$ using the perturbative QCD(PQCD)
approach. The description of final state interactions between the $D\pi$ pair is achieved through the two-meson distribution amplitudes(DAs),
which are normalized to the time-like form factor. The PQCD predictions on the branching ratios of the quasi-two-body decays $B_c\to D^*h\to D\pi h$
show an obvious  hierarchy: $Br(B_{c}^+ \to D^{*+} K^{0}\to D^0\pi^+K^{0})=({5.22}_{-0.74}^{+0.86})\times{10}^{-6},
Br(B_{c}^+ \to D^{*+} \pi^{0}\to D^0\pi^+\pi^{0})=(0.93\pm0.26)\times{10}^{-7}, Br(B_{c}^+ \to D^{*+} \eta\to D^0\pi^+\eta)
=({2.83}_{-0.52}^{+0.59})\times{10}^{-8}$ and $Br(B_{c}^+ \to D^{*+} \eta^\prime\to D^0\pi^+\eta^\prime)=({1.89}_{-0.36}^{+0.40})\times{10}^{-8}$.
 From the invariant mass $m_{D\pi}$-dependence of the decay spectrum for each channel, one can find that the branching fraction is concentrated in a narrow region
 around the $D^{*}$ pole mass.
So one can obtain the branching ratios for the corresponding two-body decays $B_c\to D^{*+}h$ under the narrow width approximation. We find that the branching ratios of
the decays $B_c\to D^{*+}h$ are consistent well with the previous PQCD calculations within errors.
These predictions will be tested by the future experiments.
\end{abstract}

{\centering\section{INTRODUCTION}\label{intro}}
In recent years, the $B_{c}$ meson decays have attracted a large amount of research interest from experimental studies \cite{shiyan1,shiyan2,shiyan3,shiyan4,shiyan5,shiyan6,shiyan7,shiyan8}.
Compared with the $B_{u,d,s}$ mesons, the $B_{c}$ meson is unique, which is composed of both heavy quarks with different flavors.
It can decay only via weak interaction, since the two flavor asymmetric quarks $(b$ and $c)$ can not annihilate into gluons (photons) via
strong (electromagnetic) interaction. The $B_{c}$ meson has many rich decay modes, because that it has sufficiently large mass and its
constituent quarks ($b$ and $c$) can decay individually. It provides a very good platform to study the nonleptonic weak decays of heavy mesons,
to test the standard model and to search for new physics signals. Along with more and more $B_{c}$ decay events being collected at the
Large Hadron Collider (LHC), the $B_{c}$ meson three-body decays will be an important research topic in both experiment and theory in
 the next few years.

At present, many approaches based on symmetry principles and factorization theorems have been used to study
the $B_{(c)}$ meson three-body decays. The former incude the U-spin \cite{sym1,sym4,sym6}, isospin and flavor $SU(3)$ symmetry
\cite{sym5,sym2,sym3},  factorization assisted topological diagram amplitude approach \cite{fat3}. The latter include the QCD-improved factorization approach \cite{qcdf1,qcdf2,qcdf5,qcdf6} and
the PQCD approach\cite{pqcd1,pqcd2,pqcd3,pqcd4,pqcd5,pqcd6,pqcd7,pqcd8,pqcd9,pqcd10,pqcd11},  where it has been proposed that the factorization theorem of three-body B decays is approximately valid when the two particles move
collinearly and the bachelor particle recoils back in the final states. Based on this quasi-two-body-decay mechanism, the two-hadron distribution
amplitudes (DAs) are introduced into the PQCD approach, where the strong dynamics between the two final hadrons in the resonant regions are included.

Previously, the corresponding two-body decays $B_{c}\to D^{*}h$ with $h=(K^0,\pi^0,\eta,\eta^{\prime})$ have been studied by the
different theories, such as the PQCD approach \cite{canshu1}, the relativistic constituent quark model (RCQM) \cite{rcqm1}.
In this work, assuming $D^{*}$ is
an internal resonance state, which further decays into $D^0\pi^+$, we will study the quasi-two-body decays  $B_c\to D^*h\to D^{0}\pi^{+}h$
using the PQCD approach. After being introduced the new non-perturbative inputs, the two-meson $D\pi$ distribution amplitudes,
the factorization formulae for the $B_c\to D^*h\to D^{0}\pi^{+}h$ decay amplitudes can be written as\cite{pqcd1,cff1,cff2}
\be
\mathcal{M}=\Phi_{B_c} \otimes H \otimes \Phi_{D\pi} \otimes \Phi_{h},
\en
where $\Phi_{B_c}(\Phi_{h})$ denotes the DAs of the initial (final bachelor) meson, $\Phi_{D\pi}$ is the $D\pi$ two-meson DAs, and
$\otimes$ means the convolution integrations over the parton momenta. The evolution of the hard kernel H for the b quark decay, similar to the two-body decay case,
 starts with the diagrams of single hard gluon exchange.

The overall layout of this paper is as follows. In section \ref{framework}, the kinematic variables for the $B_{c}$ meson three-body
decays are defined. The considered two-meson ($D\pi$) P-wave DAs are parametrized, whose normalization form factors are assumed to
take the relativistic Breit-Wigner (RBW) model. Then the Feynman diagrams and the total amplitudes for these decays are given.
In section \ref{results}, the numerical results are presented and discussed. The analytic formulas of the decay amplitudes for each Feynman
diagram are collected in section \ref{appendix} Appendix.

{\centering\section{THE FRAMEWORK}\label{framework}}
{\centering\subsection{Distribution amplitudes}}

We begin with the parametrization of the kinematic variables involved in the decays $B_c\to D^*h\to D\pi h$ with $
 h= (K^0, \pi^0, \eta, \eta^{\prime})$. In the rest frame of the $B_c$ meson,
we define the $B_c$ meson momentum $P_{B_c}$, the $D^*$ meson momentum $P$
and the bachelor meson $h$ momentum $P_3$ in the light-cone coordinates as
\be
P_{B_c} &=& \frac{m_{B_c}}{\sqrt{2}}(1,1,0_T), \quad P = \frac{m_{B_c}}{\sqrt{2}}(1,\eta,0_T), \quad P_3 = \frac{m_{B_c}}{\sqrt{2}}(0,1-\eta,0_T).
\en
with the $B_c$ meson mass $m_{B_c}$. The variable $\eta = \omega^2/m_{B_c}^2 = s/m_{B_c}^2$ with $\omega^2=s=m^2_{D\pi}$ being the square of the invariant mass of the $D\pi$ pair.
The momenta of the light quarks in the $B_c$, $D^*$ and the bachelor meson $h$ as $k_1$, $k_2$ and $k_3$, respectively,
\be
k_1 = (0,\frac{m_{B_c}}{\sqrt{2}}x_1,k_{1T}), \;\; k_2 = (\frac{m_{B_c}}{\sqrt{2}}z,0,k_{2T}), \;\; k_3 = (0,\frac{m_{B_c}}{\sqrt{2}}(1-\eta)x_3,k_{3T}),
\en
where $x_1$, $z$ and $x_3$ are the momentum fractions.

The P-wave $D\pi$ two-meson DAs are defined in the same way as Ref.\cite{DstADs1}
\be
\Phi_{D \pi}^{P-wave}=\frac{1}{\sqrt{2 N_{c}}} \not \epsilon_{L} (\not p+\sqrt{s}) \phi_{D \pi}(z, b, s),
\en
with the distribution amplitude
\be
\phi_{D \pi}(z, b, s)=\frac{F_{D \pi}(s)}{2 \sqrt{2 N_{c}}}6 z(1-z)\left[1+a_{D \pi}(1-2 z)\right] \exp \left(-\omega_{D \pi}^{2} b^{2} / 2\right),
\en
where the Gegenbauer moment $a_{D\pi} = 0.50 \pm 0.10$ and the shape parameter $\omega_{D\pi} = 0.10 \pm 0.02 $ GeV \cite{DstADs1}.

The strong interactions between the resonance and the final-state meson pair can be factorized into the time-like
form factor, which is guaranteed by the Watson theorem\cite{kw}. For the narrow resonances, the RBW function \cite{rbw} is
a convenient model to well separate from any other resonant or nonresonant contributions with
the same spin, and has been widely used in the experimental data analyses. Here, the time-like form factor $F_{D\pi}(s)$
can be defined through the matrix element
\be
\langle D(p_1)\pi(p_2)|\bar c\gamma_\mu(1-\gamma_5)q|0\rangle=\left[(p_1-p_2)_\mu-\frac{m^2_D-m^2_\pi}{p^2}p_\mu\right]F_{D\pi}(s)
+\frac{m^2_D-m^2_\pi}{p^2}p_\mu F_0(s),
\en
where $p=p_1+p_2$, $p_1(p_2)$ and $m_D(m_\pi)$ are the $D(\pi)$ meson momentum and mass, respectively. $F_{D\pi}(s)$ and $F_0(s)$
are the P-wave and S-wave form factors for the $D\pi$ system.  $F_{D\pi}(s)$
is parameterized with the RBW line shape
\be
F_{D \pi}(s)=\frac{\sqrt{s} f_{D^*} g_{D^{*}D \pi}}{m_{D^{*}}^{2}-s-i m_{D^{*}} \Gamma(s)},
\en
where $f_{D^*}$ and $m_{D^{*}}$ are the decay constant and the pole mass of the $D^*$ meson, respectively. The coupling constant $g_{D^{*}D \pi}$ can
be determined through the decay width $\Gamma(s)$.  The invariant mass dependent decay
width $\Gamma(s)$ is defined as
\be
\Gamma(s)= \Gamma_{0} \Big(\frac{q}{q_{0}} \Big)^3 \Big(\frac{m_{D^{*}}}{\sqrt{s}}\Big)\emph{X}^{2}(q r_{BW}),
\en
where the Blatt-Weisskopf barrier factor \cite{blatt}
\be
\emph{X}(q r_{BW}) = \sqrt{\frac{1+(q_{0} r_{BW})^{2}}{1+(q r_{BW})^{2}}},
\en
with the barrier radius $r_{BW} = 4.0 GeV^{-1}$ Refs.\cite{rbw2,rbw3}, $q$ being the momentum for the daughter meson $D$ or $\pi$ in the
$D^*$ meson rest frame
\be
q=\frac{1}{2} \sqrt{\left[s-\left(m_{D}+m_{\pi}\right)^{2}\right]\left[s-\left(m_{D}-m_{\pi}\right)^{2}\right] / s},
\en
and $q_0$ being the value of $q$ at $s = m_{D^{*}}^2$.

The twist-2 distribution amplitude $\phi_{h}^{A}$, and the twist-3 ones $\phi_{h}^{P}$ and $\phi_{h}^{T}$
have been parameterized as\cite{PADs1,PADs2,PADs3}
\be
\phi_{h}^{A}(x) &=& \frac{f_{h}}{2 \sqrt{2 N_{c}}} 6 x (1-x)\Big[1+a_{1}^{h} C_{1}^{3/2} (2 x-1)+a_{2}^{h}  C_{2}^{3/2} (2 x-1)+a_{4}^{h} C_{4}^{3/2} (2 x-1) \Big],\\
\phi_{h}^{P}(x) &=& \frac{f_{h}}{2 \sqrt{2 N_{c}}} \Big[1+\left( 30 \eta_{3}-\frac{5}{2} \rho_{h}^{2} \right) C_{2}^{1/2} (2 x-1)-3 \left( \eta_{3} \omega_{3}+\frac{9}{20} \rho_{h}^{2}(1+6 a_{2}^{h}) \right) C_{4}^{1/2} (2 x-1) \Big],\\
\phi_{h}^{T}(x) &=& \frac{f_{h}}{2 \sqrt{2 N_{c}}}(1-2 x)\Big[1+6 \left( 5 \eta_{3}-\frac{1}{2} \eta_{3} \omega_{3}-\frac{7}{20} \rho_{h}^{2}-\frac{3}{5}\rho_{h}^{2} a_{2}^{P} \right)(1-10 x+10 x^{2}) \Big],
\en
where the subscript $h$ represents the pseudoscalar mesons $K, \pi$ and the flavor states $\eta_q=\frac{u\bar u+d\bar d}{\sqrt2},
\eta_s=s\bar s$.
The parameters
$\eta_{3}=0.015, \omega_{3}=-3$, the mass ratios $\rho_{K(\pi)} = m_{K(\pi)}/m_{0K(\pi)}, \rho_{\eta_q}=2m_q/m_{qq},
\rho_{\eta_s}=2m_s/m_{ss}$ with $m_{0K(\pi)}, m_{qq}$ and $m_{ss}$ being the chiral enhancemnet scales.
The Gegenbauer polynomials $C_{n}^{\nu}(t)$
\be
C_{2}^{1/2}(t) &=& \frac{1}{2}(3 t^{2}-1),\quad C_{4}^{1/2}(t) = \frac{1}{8}(3-30 t^{2}+35 t^{4}),\\
C_{1}^{3/2}(t) &=& 3 t, \quad C_{2}^{3/2}(t) = \frac{3}{2}(5 t^{2}-1),\quad C_{4}^{3/2}(t) = \frac{15}{8}(1-14 t^{2}+21 t^{4}).
\en
The parameters of the hadronic wave functions are taken from
Refs. \cite{hua,pball}
\be
a_{1}^{K}&=&-0.108\pm0.053, \quad a_{2}^{K}=0.170\pm0.046 \quad a_{4}^{K}=0.073\pm0.022,  \quad a_{1}^{\pi}=0, \quad
a_{2}^{\pi}=0.258\pm0.087,  \\
\quad a_{4}^{\pi}&=&0.122\pm0.055,\quad a_{1}^{\eta_{q}}=a_{1}^{\eta_{s}}=0,\quad  a_{2}^{\eta_{q}}=a_{2}^{\eta_{s}}=0.115\pm0.115, \quad a_{4}^{\eta_{q}}=a_{4}^{\eta_{s}}=-0.015.
\en

For the wave function of the heavy $B_c$ meson, we take as
\be\label{eq8}
\int d^{4} z e^{i k_{1} \cdot z}\left\langle 0\left|\bar{b}_{\alpha}(0) c_{\beta}(z)\right| B_{c}\left(P_{B_c}\right)\right\rangle=\frac{i}{\sqrt{2 N_{c}}}\left[\left(P\hspace{-2.3truemm}/_{B_c}+m_{B_c}\right) \gamma_{5} \phi_{B_{c}}\left(k_{1}\right)\right]_{\beta \alpha},
\en
where we only consider the contribution from the dominant Lorentz structure. In coordinate space
the distribution amplitude $\phi_{B_c}$ with an intrinsic $b$ (the conjugate space coordinate to $k_T$ ) dependence is adopted in a
Gaussian form as \cite{xliu}
\be
\phi_{B_{c}}(x, b)=N_{B_{c}} x(1-x) \exp \left[-\frac{(1-x) m_{c}^{2}+x m_{b}^{2}}{8 \omega_b^{2} x(1-x)}-2 \omega_b^{2} b^{2} x(1-x)\right],
\en
where the shape parameter  $\omega_b = 1.0 \pm 0.1 \mathrm{GeV}$  related to the factor  $N_{B_{c}}$  by the normalization
$\int_{0}^{1} \phi_{B_{c}}(x, 0) d x=1$.
{\centering\subsection{Analytic formulae}}

For the quasi-two-body decays $B_c\to D^*h\to D\pi h$, the effective Hamiltonian relevant to the $b\to D (D = d,s)$ transition is given by Ref. \cite{heff2}
\be
\bea
H_{eff} = &\frac{G_F}{\sqrt{2}}[\sum_{q=u,c}V_{qb}V_{qD}^*\{C_1(\mu)O_1^{(q)}(\mu)+C_2(\mu)O_2^{(q)}(\mu)\}
-\sum_{i=3\sim10}V_{tb}V_{tD}^*C_i(\mu)O_i(\mu)],
\ena
\en
where the Fermi coupling constant $G_F\simeq 1.166\times 10^{-5}GeV^{-2}$ \cite{pdg}, $V_{qb}V_{qD}^*$ and $V_{tb}V_{tD}^*$
are the products of the Cabibbo-Kobayashi-Maskawa (CKM) matrix elements. The scale $\mu$ separates the effective Hamiltonian into two distinct parts: the Wilson
coefficients $C_i$ and the local four-quark operators $O_i$. The local four-quark operators are written as
\be{}
\bea{}
O_{1}^{(q)} & =  \left(\bar{D}_{i} q_{j}\right)_{V-A}\left(\bar{q}_{j} b_{i}\right)_{V-A},  &O_{2}^{(q)} =&  \left(\bar{D}_{i} q_{i}\right)_{V-A}\left(\bar{q}_{j} b_{j}\right)_{V-A}, \\
O_{3} & =  \left(\bar{D}_{i} b_{i}\right)_{V-A} \sum_{q^{\prime}}\left(\bar{q}_{j}^{\prime} q_{j}^{\prime}\right)_{V-A}, & O_{4} =&  \left(\bar{D}_{i} b_{j}\right)_{V-A} \sum_{q^{\prime}}\left(\bar{q}_{j}^{\prime} q_{i}^{\prime}\right)_{V-A}, \\
O_{5} & =  \left(\bar{D}_{i} b_{i}\right)_{V-A} \sum_{q^{\prime}}\left(\bar{q}_{j}^{\prime} q_{j}^{\prime}\right)_{V+A}, & O_{6} =&  \left(\bar{D}_{i} b_{j}\right)_{V-A} \sum_{q^{\prime}}\left(\bar{q}_{j}^{\prime} q_{i}^{\prime}\right)_{V+A}, \\
O_{7} & =  \frac{3}{2}\left(\bar{D}_{i} b_{i}\right)_{V-A} \sum_{q^{\prime}} e_{q^{\prime}}\left(\bar{q}_{j}^{\prime} q_{j}^{\prime}\right)_{V+A},& O_{8} =&  \frac{3}{2}\left(\bar{D}_{i} b_{j}\right)_{V-A} \sum_{q^{\prime}} e_{q^{\prime}}\left(\bar{q}_{j}^{\prime} q_{i}^{\prime}\right)_{V+A}, \\
O_{9} & =  \frac{3}{2}\left(\bar{D}_{i} b_{i}\right)_{V-A} \sum_{q^{\prime}} e_{q^{\prime}}\left(\bar{q}_{j}^{\prime} q_{j}^{\prime}\right)_{V-A},& O_{10} =& \frac{3}{2}\left(\bar{D}_{i} b_{j}\right)_{V-A} \sum_{q^{\prime}} e_{q^{\prime}}\left(\bar{q}_{j}^{\prime} q_{i}^{\prime}\right)_{V-A},
\ena{}
\en{}
with the color indices $i$ and $j$. Here $V\pm A$ refer to the Lorentz structures $\gamma_\mu(1\pm \gamma_5)$.

\begin{figure} [htbp]
\centering
\begin{minipage}[t]{0.45\linewidth}
\includegraphics[scale=0.45]{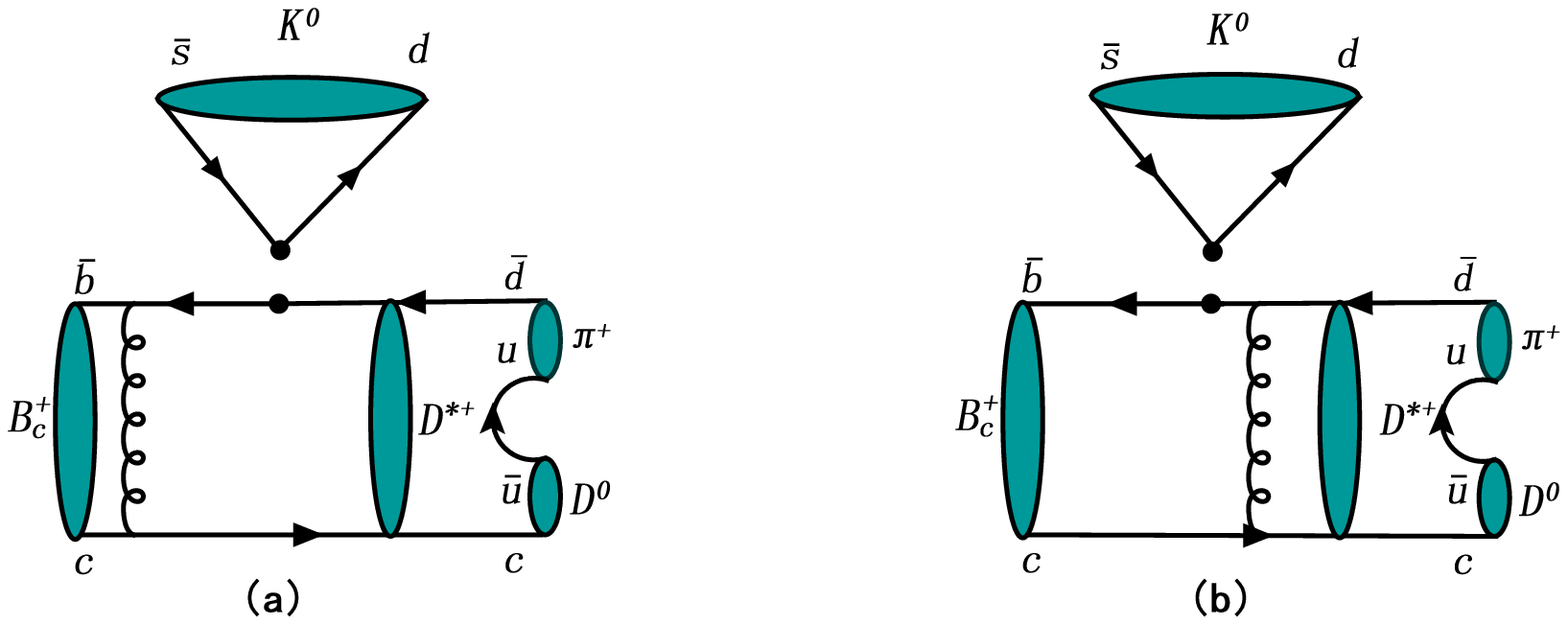}
\includegraphics[scale=0.45]{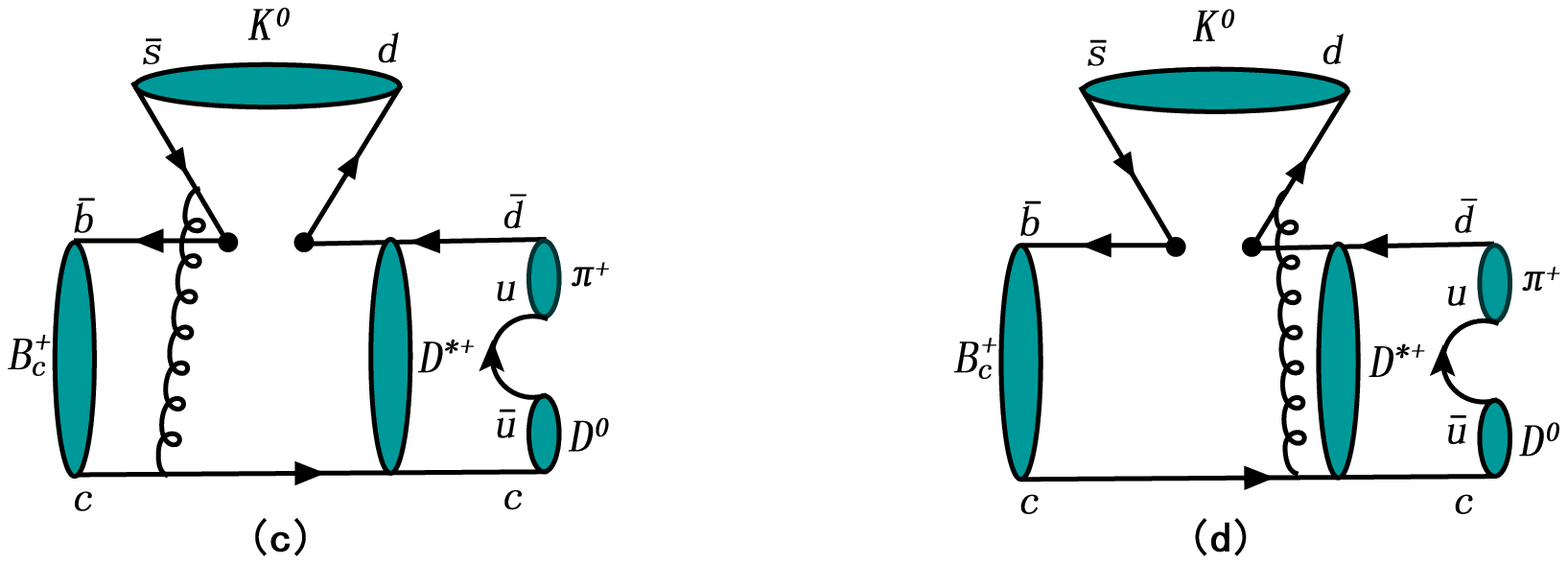}
\caption{The factorizable (a, b) and non-factorizable (c, d) emission diagrams for the decay
 $B_c^+\to D^{*+}K^{0} \to D^{0}\pi^{+} K^{0}$. }\label{img1}
\end{minipage}
\hspace{8mm}
\begin{minipage}[t]{0.45\linewidth}
\includegraphics[scale=0.5]{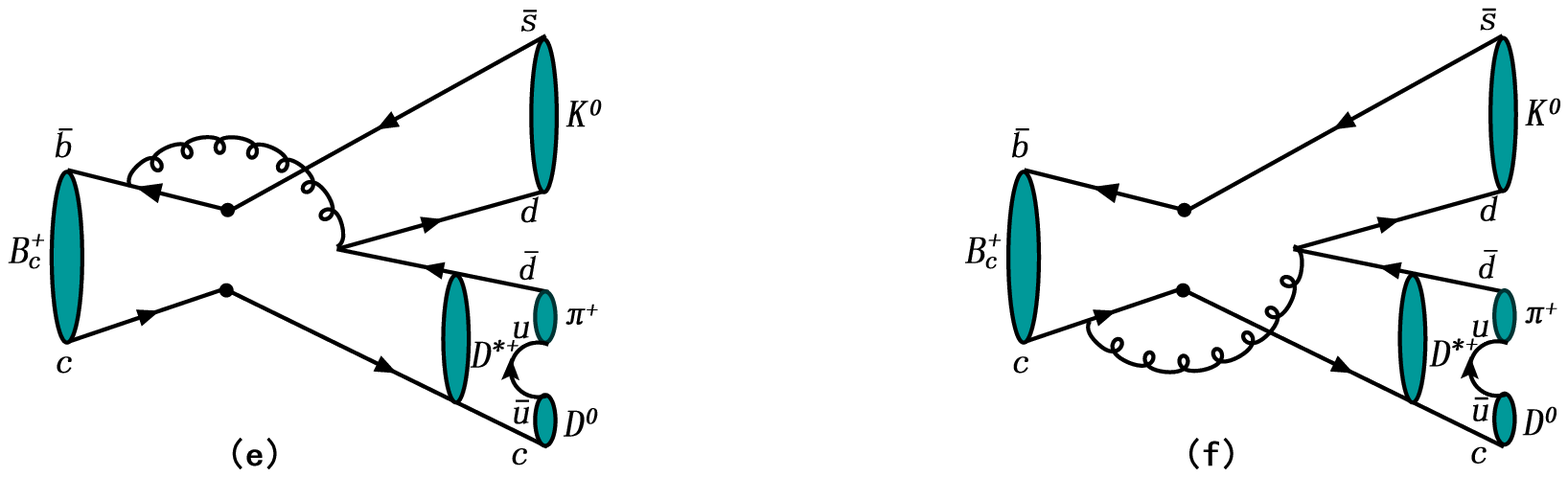}
\includegraphics[scale=0.5]{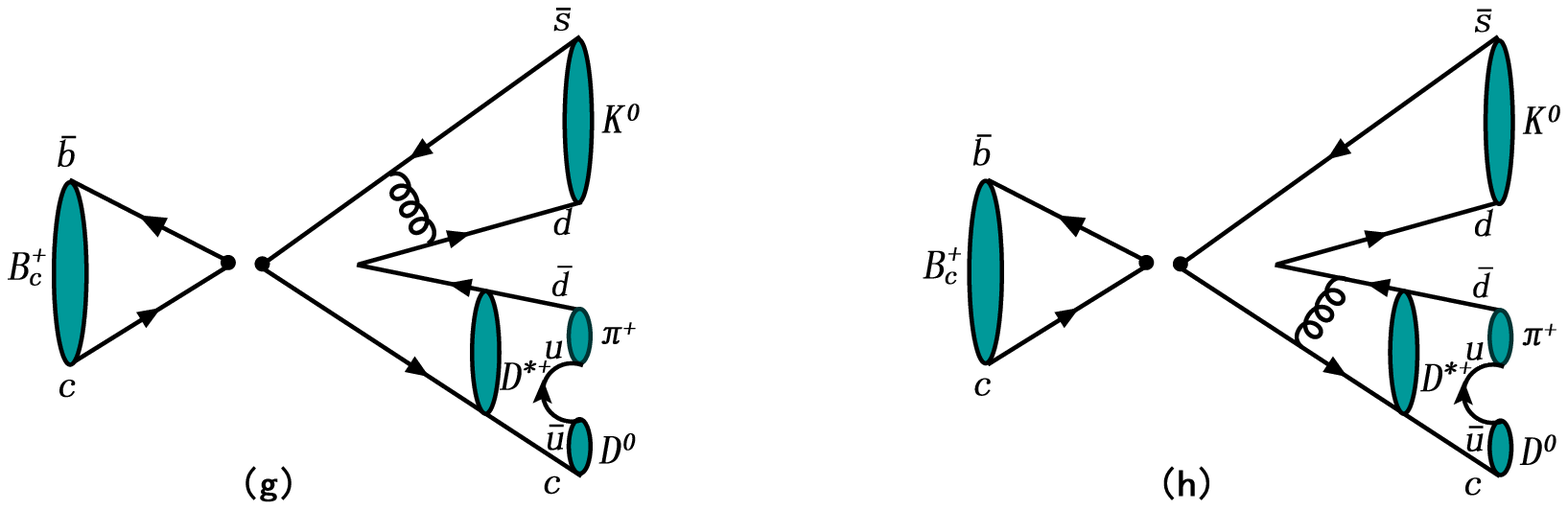}
\vspace{-3mm}
\caption{The nonfactorizable (e, f) and factorizable (g, h) annihilation diagrams for the decay
$B_c^+\to D^{*+}K^{0} \to D^{0}\pi^{+} K^{0}$. }\label{img2}
\end{minipage}
\end{figure}

The typical Feynman diagrams at the leading order for the quasi-two-body
decays $B_c^+\to D^{*+}h\to D^0\pi^+ h$ are shown in Fig. \ref{img1} and Fig. \ref{img2}, where we take the decay
$B_c^+\to D^{*+}K^{0} \to D^{0}\pi^{+} K^{0}$ as an example. The analysis formulas for the decay amplitudes of each Feynman diagram are given in the appendix.

By combining the amplitudes from the different
Feynman diagrams, the total decay amplitudes for these considered decays are given as
\be\label{eq11}
\mathcal{A}\left(B_{c}^+ \rightarrow D^{*+} K^{0}\rightarrow D^0\pi^+K^{0}\right)&=& V_{cs}V^*_{cb}\left[a_{1} \mathcal{F}_{a}^{L L}+C_{1} \mathcal{M}_{a}^{L L}\right]-V_{ts}V^*_{tb}\left[\left(C_{3}-\frac{1}{2} C_{9}\right) \mathcal{M}_{e}^{L L}\right.\non
&&+\left(C_{3}+C_{9}\right) \mathcal{M}_{a}^{L L}+\left(C_{5}-\frac{1}{2} C_{7}\right) \mathcal{M}_{e}^{L R}+\left(C_{5}+C_{7}\right) \mathcal{M}_{a}^{L R} \non
&&+\left(C_{4}+\frac{1}{3} C_{3}+C_{10}+\frac{1}{3} C_{9}\right) \mathcal{F}_{a}^{L L}+\left(C_{4}+\frac{1}{3} C_{3}-\frac{1}{2} C_{10}\right.\non
&&\left.-\frac{1}{6} C_{9}\right) \mathcal{F}_{e}^{L L}+(C_{6}+\frac{1}{3} C_{5}-\frac{1}{2} C_{8}-\frac{1}{6} C_{7}) \mathcal{F}_{e}^{S P}\non
&&\left.+\left(C_{6}+\frac{1}{3} C_{5}+C_{8}+\frac{1}{3} C_{7}\right) \mathcal{F}_{a}^{S P}\right],
\en
\be\label{eq12}
\sqrt{2} \mathcal{A}\left(B_{c}^+ \rightarrow D^{*+} \pi^{0}\rightarrow D^0\pi^+ \pi^{0}\right)&=& V_{ud}V^*_{ub}\left[a_{2} \mathcal{F}_{e}^{L L}+C_{2} \mathcal{M}_{e}^{L L}\right]
-V_{cd}V^*_{cb}\left[a_{1}\mathcal{F}_{a}^{L L}+C_{1} \mathcal{M}_{a}^{L L}\right]\non
&&-V_{td}V^*_{tb}\left[\left(\frac{3}{2} C_{10}-C_{3}+\frac{1}{2} C_{9}\right) \mathcal{M}_{e}^{L L}-\left(C_{3}+C_{9}\right) \mathcal{M}_{a}^{L L}\right.\non
&&+\left(-C_{5}+\frac{1}{2} C_{7}\right) \mathcal{M}_{e}^{L R}+\left(-C_{4}-\frac{1}{3} C_{3}-C_{10}-\frac{1}{3} C_{9}\right) \mathcal{F}_{a}^{L L}\non
&&+\left(C_{10}+\frac{5}{3} C_{9}-\frac{1}{3} C_{3}-C_{4}-\frac{3}{2} C_{7}-\frac{1}{2} C_{8}\right) \mathcal{F}_{e}^{L L} \non
&&+\left(-C_{6}-\frac{1}{3} C_{5}+\frac{1}{2} C_{8}+\frac{1}{6} C_{7}\right) \mathcal{F}_{e}^{S P}-\left(C_{5}+C_{7}\right) \mathcal{M}_{a}^{L R} \non
&&\left.+\left(-C_{6}-\frac{1}{3} C_{5}-C_{8}-\frac{1}{3} C_{7}\right) \mathcal{F}_{a}^{S P}\right],
\en
\be\label{eq13}
\sqrt{2} \mathcal{A}\left(B_{c}^+ \rightarrow D^{*+} \eta_{q}\rightarrow D^0\pi^+ \eta_{q}\right) &= &V_{ud}V^*_{ub}\left[a_{2} \mathcal{F}_{e}^{L L}+C_{2} \mathcal{M}_{e}^{L L}\right]+V_{cd}V^*_{cb}\left[a_{1} \mathcal{F}_{a}^{L L}+C_{1} \mathcal{M}_{a}^{L L}\right] \non
&& -V_{td}V^*_{tb}\left[\left(2 C_{4}+C_{3}+\frac{1}{2} C_{10}-\frac{1}{2} C_{9}\right) \mathcal{M}_{e}^{L L}+\left(C_{3}+C_{9}\right) \mathcal{M}_{a}^{L L}\right. \non
&& +\left(C_{5}-\frac{1}{2} C_{7}\right) \mathcal{M}_{e}^{L R}+\left(C_{5}+C_{7}\right) \mathcal{M}_{a}^{L R}+\left(C_{4}+\frac{1}{3} C_{3}+C_{10}\right. \non
&& \left.+\frac{1}{3} C_{9}\right) \mathcal{F}_{a}^{L L}+\left(\frac{7}{3} C_{3}+\frac{5}{3} C_{4}+\frac{1}{3}\left(C_{9}-C_{10}\right)\right) \mathcal{F}_{e}^{L L}+\left(2 C_{5}\right. \non
&& \left.+\frac{2}{3} C_{6}+\frac{1}{2} C_{7}+\frac{1}{6} C_{8}\right) \mathcal{F}_{e}^{L R}+\left(C_{6}+\frac{1}{3} C_{5}-\frac{1}{2} C_{8}-\frac{1}{6} C_{7}\right) \mathcal{F}_{e}^{S P} \non
&& \left.+\left(C_{6}+\frac{1}{3} C_{5}+C_{8}+\frac{1}{3} C_{7}\right) \mathcal{F}_{a}^{S P}\right],
\en
\be\label{eq14}
\mathcal{A}\left(B_{c}^+ \rightarrow D^{*+} \eta_{s}\rightarrow D^0\pi^+ \eta_{s}\right) &= & -V_{td}V^*_{tb}\left[\left(C_{4}-\frac{1}{2} C_{10}\right) \mathcal{M}_{e}^{L L}+\left(C_{6}-\frac{1}{2} C_{8}\right) \mathcal{M}_{e}^{S P}+\left(C_{3}+\frac{1}{3} C_{4}\right.\right. \non
&& \left.\left.-\frac{1}{2} C_{9}-\frac{1}{6} C_{10}\right) \mathcal{F}_{e}^{L L}+\left(C_{5}+\frac{1}{3} C_{6}-\frac{1}{2} C_{7}-\frac{1}{6} C_{8}\right) \mathcal{F}_{e}^{L R}\right],
\en
where the combinations of the Wilson coefficients  $a_{1}=C_{2}+C_{1} / 3$  and $a_{2}=C_{1}+C_{2} / 3 $.

It should be noticed that,  Eq.(\ref{eq13}) and Eq.(\ref{eq14}) are the decay amplitudes corresponding to the flavor states
 $\eta_{q}$ and $\eta_{s}$, respectively. For the physical states $(\eta, \eta^{\prime})$, the decay amplitudes are written as
\be
\mathcal{A}\left(B_{c}^+ \rightarrow D^{*+} \eta\rightarrow D^0\pi^+ \eta\right) &=& \mathcal{A}\left(B_{c}^+ \rightarrow D^{*+} \eta_{q}\rightarrow D^0\pi^+ \eta_{q}\right) \cos \phi-\mathcal{A}\left(B_{c}^+ \rightarrow D^{*+} \eta_{s}\rightarrow D^0\pi^+ \eta_{s}\right) \sin \phi, \\
\mathcal{A}\left(B_{c}^+ \rightarrow D^{*+} \eta^{\prime}\rightarrow D^0\pi^+ \eta^{\prime}\right) &=& \mathcal{A}\left(B_{c}^+ \rightarrow D^{*+} \eta_{q}\rightarrow D^0\pi^+ \eta_{q}\right) \sin \phi+\mathcal{A}\left(B_{c}^+ \rightarrow D^{*+} \eta_{s}\rightarrow D^0\pi^+ \eta_{s}\right) \cos \phi,\label{etap}
\en
where $\phi=39.3^{\circ}\pm1.0^\circ$ \cite{feld} is the mixing angle between these two flavor states and defined as
\be
\left(\begin{array}{c}
\eta \\
\eta^{\prime}
\end{array}\right) & = \left(\begin{array}{cc}
\cos \phi & -\sin \phi \\
\sin \phi & \cos \phi
\end{array}\right)\left(\begin{array}{l}
\eta_{q} \\
\eta_{s}
\end{array}\right).
\en

Then the differential decay rate is described as\cite{br1,br2}
\be\label{eq15}
\frac{d \mathcal{B}}{d s}=\frac{\tau_{B_c} q^{3} q^{3}_{h}}{48 \pi^{3} m_{B_c}^{7}} \overline{|{\mathcal{A}}|^{2}},
\en
where $\tau_{B_c}$ is the $B_c$ meson mean lifetime, the kinematic variable $q_h$ denotes the momentum magnitude of
the bachelor meson $h$ in the center-of-mass frame of the $D\pi$ pair,
\be
q_{h}=\frac{1}{2}\sqrt{[(m_{B_c}^2-m_h^2)^2-2(m_{B_c}^2+m_h^2)s+s^2]/s}.
\en
{\centering\section{Numerical results}\label{results}}

   The adopted input parameters in our numerical calculations are summarized as following (the masses,
decay constants and QCD scale are in units of GeV, the $B_c$ meson lifetime is in units of ps)\cite{canshu1,PADs3,pdg}:
\be
\Lambda^{(5)}_{QCD} &=& 0.112,\; m_{B_c^{+}} = 6.27447,\; m_b = 4.18,\; m_{K^0} = 0.498,\; m_{0K} = 1.7,\; m_{\pi^{\pm}} = 0.140, \\
 m_{\pi^0}&=& 0.135,\; m_{0\pi}= 1.3,\;m_\eta=0.548,\; m_{\eta^\prime}=0.958,\; m_{qq}=0.110, \;m_{ss}=0.707,  \\
 m_{D^{*\pm}} &=& 2.010,\;m_{D^{0}}=1.865,\;\tau_{B_{c}^+}= 0.510,\; f_{B_c} = 0.489,\; f_{D^*} = 0.25,\;f_{K} = 0.16, \\
f_{\pi}  &=& 0.13,\;f_{\eta_{q}}=(1.07\pm0.02)f_{\pi},\; f_{\eta_{s}} = (1.34\pm0.06)f_{\pi}.
\en
As to the CKM matrix elements, we employ the Wolfenstein parametrization with the inputs\cite{pdg}
\be
\lambda&=&0.22453 \pm 0.00044, \quad A=0.836 \pm 0.015,\\
\bar{\rho}&=&0.122_{-0.017}^{+0.018}, \quad \bar{\eta}=0.355_{-0.011}^{+0.012}.
\en

Using the decay amplitudes as given in the Appendix,
the total amplitudes listed in Eq.(\ref{eq12}) - Eq.(\ref{etap}) and the differential branching ratio shown in Eq. (\ref{eq15}), integrating over the full $D\pi$ invariant mass region
 $(m_D + m_{\pi})\leq \sqrt{s}\leq (m_{B_c}-m_h)^2$ with $h=(K^0, \pi^0, \eta^{(\prime)})$, we obtain the branching ratios for the quasi-two-body decays
\be\label{bran}
Br\left(B_{c}^+ \to D^{*+} K^{0}\rightarrow D^0\pi^+K^{0}\right) &= \left({5.22}_{-0.38-0.45-0.00-0.21-0.15-0.37}^{+0.53+0.47+0.21+0.09+0.15+0.41}\right)\times{10}^{-6},\\
Br\left(B_{c}^+ \rightarrow D^{*+} \pi^{0}\rightarrow D^0\pi^+ \pi^{0} \right) &= \left({0.93}_{-0.08-0.08-0.23-0.03-0.02-0.05}^{+0.11+0.09+0.21+0.01+0.03+0.06}\right)\times{10}^{-7},\\
Br\left(B_{c}^+ \rightarrow D^{*+} \eta\rightarrow D^0\pi^+\eta \right) &= \left({2.83}_{-0.19-0.23-0.21-0.25-0.20-0.18}^{+0.22+0.25+0.19+0.28+0.18+0.31}\right)\times{10}^{-8},\\
Br (B_{c}^+ \rightarrow D^{*+} \eta^{\prime}\rightarrow D^0\pi^+\eta^{\prime} ) &= \left({1.89}_{-0.12-0.16-0.14-0.15-0.16-0.15}^{+0.13+0.17+0.12+0.19+0.14+0.21}\right)\times{10}^{-8},
\en
where the first error originates from the shape parameter in the $B_c$ meson DAs, $\omega_{B_c}=1.0\pm0.1$ GeV,
the second one comes from the decay constant $f_{D^{*}}=(250\pm11)$ MeV,
the third and the forth errors
are induced by the Gegenbauer coefficient
 $a_{D\pi}=0.50\pm0.10$ and the shape parameter $\omega_{D\pi}=0.10\pm0.02$ GeV in the $D\pi$ pair DAs, respectively,
 the fifth one is caused by the decay width
of the resonance $D^{*+}$, $\Gamma_{D^{*+}}=83.3\pm2.6$ keV, the last one is induced by
the next-to-leading-order effect in the PQCD approach by changing the hard scale $t$ from 0.75t to 1.25t.
As to other errors, which come from the uncertainties of the parameters in the DAs of
the bachelor meson $h$, the Wolfenstein parameters etc., have been neglected since they are very small.

If we assume the isospin conservation for the strong
decay $D^*\to D\pi$,
\be\label{20}
\frac{\Gamma(D^{*+}\to D^0\pi^+)}{\Gamma(D^{*+}\to D\pi)}=2/3.
\en
Under the narrow width approximation relation, the branching ratios of these quasi-two-body decays can be written as
\be\label{21}
Br\left(B_{c}^+\to D^{*+} h \to D^0\pi^+ h \right) = Br(B_{c}^+\to D^{*+}h)\cdot Br(D^{*+}\to D^0\pi^+).
\en
Based on Eq. (\ref{20}) and Eq. (\ref{21}), we can obtain the branching ratios of the corresponding two-body decays
\be\label{22}
Br(B_{c}^+\to D^{*+}K^0) &= \left({7.83}_{-0.57-0.67-0.00-0.31-0.23-0.55}^{+0.79+0.70+0.32+0.13+0.22+0.61}\right)\times{10}^{-6},\\
Br(B_{c}^+\to D^{*+}\pi^0) &= \left({1.40}_{-0.12-0.12-0.35-0.05-0.03-0.07}^{+0.16+0.13+0.32+0.01+0.04+0.09}\right)\times{10}^{-7},\\
Br(B_{c}^+\to D^{*+}\eta) &= \left({4.25}_{-0.29-0.35-0.32-0.38-0.30-0.27}^{+0.33+0.38+0.29+0.42+0.27+0.47}\right)\times{10}^{-8},\\
Br(B_{c}^+\to D^{*+} \eta^{\prime}) &=\left({2.84}_{-0.18-0.24-0.21-0.23-0.24-0.23}^{+0.20+0.26+0.18+0.29+0.21+0.32}\right)\times{10}^{-8},
\en
where we assume the branching ratio of the decay $D^{*+}\to D\pi$ to be $100\%$, the isospin conservation and the narrow width approximation to be observed.

The branching ratios of the two-body decays $B_c\to D^*h$ with $h= (K^0, \pi^0, \eta, \eta^{\prime})$ have been calculated in the
PQCD approach \cite{canshu1}, where the results were given as
\be\label{23}
Br(B_{c}^+\to D^{*+}K^0) &=& \left({7.78}_{-2.40-0.02-0.52}^{+2.54+0.02+0.72}\right)\times{10}^{-6},\\
Br(B_{c}^+\to D^{*+}\pi^0) &=& \left({1.3}_{-0.3-0.3-0.0}^{+0.4+0.2+0.0}\right)\times{10}^{-7},\\
Br(B_{c}^+\to D^{*+}\eta) &=& \left({3.4}_{-0.9-1.5-0.00}^{+1.4+1.9+0.4}\right)\times{10}^{-8},\\
Br(B_{c}^+\to D^{*+} \eta^{\prime}) &=&\left({1.5}_{-0.5-0.6-0.1}^{+0.8+0.8+0.3}\right)\times{10}^{-8}.
\en
From upper formulas, one can find that the branching ratios of the decays $B_{c}^+\to D^{*+}h$
calculated in both two-body and three-body frameworks are consistent with each other within errors.
The decay widths for the decays $B_{c}^+\to D^{*+}K^0$ and $B_{c}^+\to D^{*+}\pi^0$ were calculated by using
the RCQM \cite{rcqm1}, which are given as $\Gamma(B_{c}^+\to D^{*+}K^0)=4.10\times10^{-19}$ GeV and
$\Gamma(B_{c}^+\to D^{*+}\pi^0)=9.83\times10^{-20}$ GeV, respectively. If taking $\tau_{B_{c}^+}= 0.510$ ps, we can obtain their branching ratios as
$BR(B_{c}^+\to D^{*+}K^0)=3.18\times10^{-7}$ and $Br(B_{c}^+\to D^{*+}\pi^0)=0.76\times10^{-7}$, which are much smaller
than the PQCD predictions. From Eq.(\ref{eq11}) and Eq.(\ref{eq12}), one can find that the dominant contributions for these two decays come
from the factorization annihilation
amplitude $\mathcal{F}_{a}^{L L}$  asociated with the large Wilson coefficient $a_1$. Unfortunately, such kind of annihilation contribution is not
calculable under the RCQM. We hope that it can be verified by the future experiments. As we know,
the annihilation diagram contributions associated with the CKM matrix elements $V^*_{cb}V_{cs}$ are dominant in both of the
decays $B^+_c\to D^{*0}K^+$ and $B^+_c\to D^{*+}K^0$. In fact, such contributions are the same for these two channels,
so we argue that the branching ratios for these two decays $B^+_c\to D^{*0}K^+$ and $B^+_c\to D^{*+}K^0$ should be close to each other.
For example,  $Br(B^+_c\to D^{*0}K^+)=(7.35^{+3.28}_{-2.34})\times10^{-6}$ and $Br(B^+_c\to D^{*+}K^0)=(7.78^{+2.64}_{-2.46})\times10^{-6}$
were given by the PQCD approach \cite{canshu1}.
In Ref. \cite{lhcb}, the upper limit for $R_{D^{*0}K^+}$ at $95\%$ confidence level was given as
\be
R_{D^{*0}K^+}=\frac{f_c}{f_u}Br(B^+_c\to D^{*0}K^+)<1.1\times 10^{-6},
\en
where $f_c/f_u$ is the ratio of the inclusive production cross-sections of $B^+_c$ and $B^+$ mesons, which can be related the decays
$B^+_c\to J/\Psi\pi^+$ and $B^+\to J/\Psi K^+$ through the following formula
\be
R_{c/u}=\frac{f_c}{f_u}\frac{Br(B^+_c\to J/\Psi \pi^+)}{Br(B^+\to J/\Psi K^+)},
\en
which has been measured as $R_{c/u}=(0.68\pm0.10)\%$ by the LHCb collaboration \cite{lhcb3}. Unfortunately,
$Br(B^+_c\to J/\Psi\pi^+)$ has not been well measured by experiment. According to different theoretical predictions,
the LHCb collaboration gave a range $0.004\sim0.012$ for the $f_c/f_u$ values \cite{lhcb}.
Then we can obtain a lowest upper limit for the branching ratio of the decay $B^+_c\to D^{*0}K^+$
\be
Br(B^+_c\to D^{*0}K^+)<9.2\times 10^{-5}.
\en
Certainly, here the upper limit is strongly dependent on the branching ratio of the decay $B^+_c\to J/\Psi\pi^+$.
As a loosely estimate, this upper limit can also
be applied to the branching ratio of the decay $B^+_c\to D^{*+}K^0$.  Our prediction is found to satisfy this limit, which can be tested by the present LHCb experiments.
There exists constructive (destrutive) interference between the amplitudes
$\mathcal{A}\left(B_{c}^+ \to D^{*+} \eta_{q}\to D^0\pi^+ \eta_{q}\right)$ and
$\mathcal{A} (B_{c}^+\to D^{*+} \eta_{s}\to D^0\pi^+ \eta_{s})$ in the decay
$B_{c}^+\to D^{*+}\eta\to D^0\pi^+ \eta (B_{c}^+\to D^{*+}\eta^\prime\to D^0\pi^+ \eta^\prime)$, which
enhances (reduces) the branching ratio of the corresponding decay.

\begin{figure}[htbp]
\centering
\includegraphics[scale=0.3]{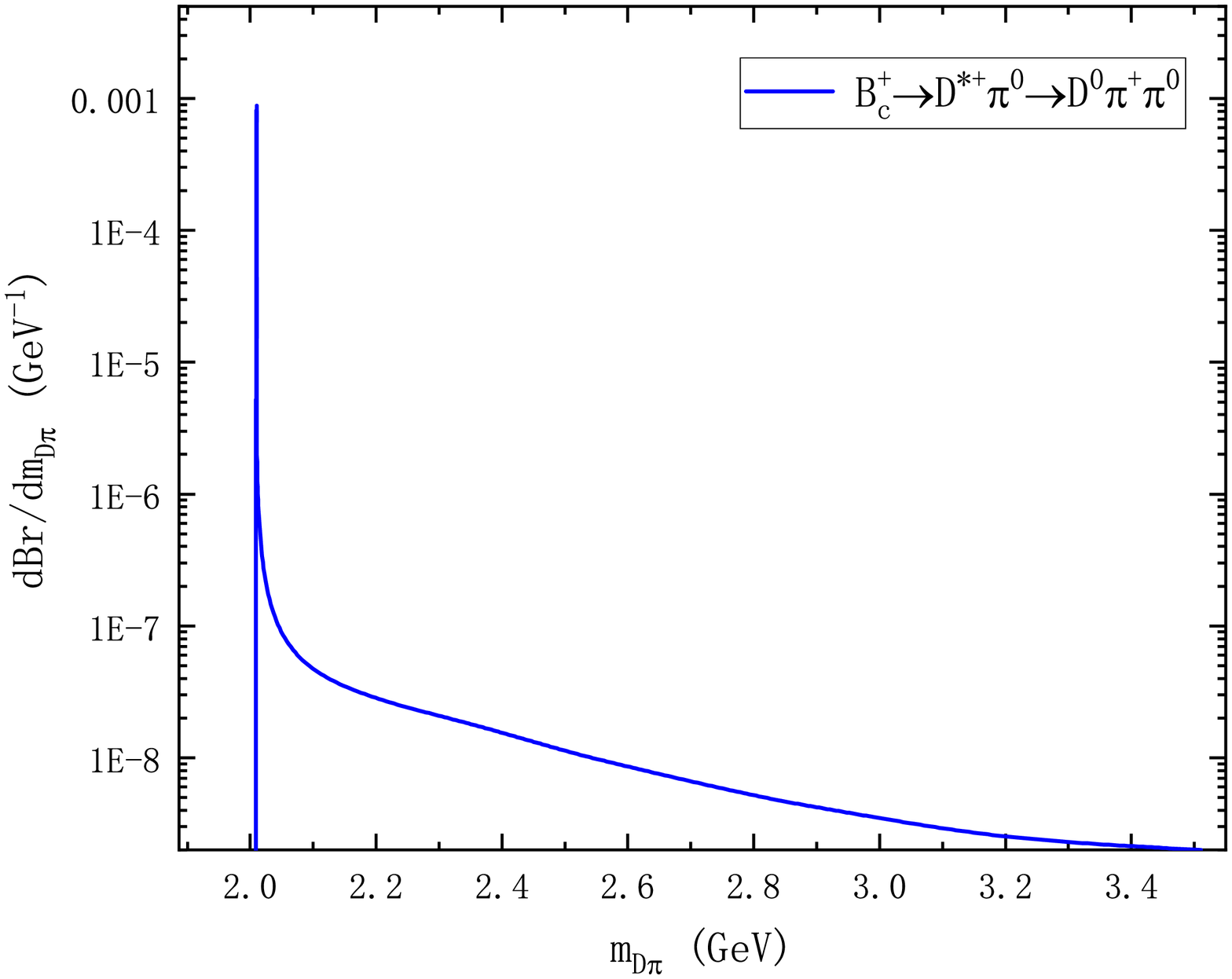}
\includegraphics[scale=0.3]{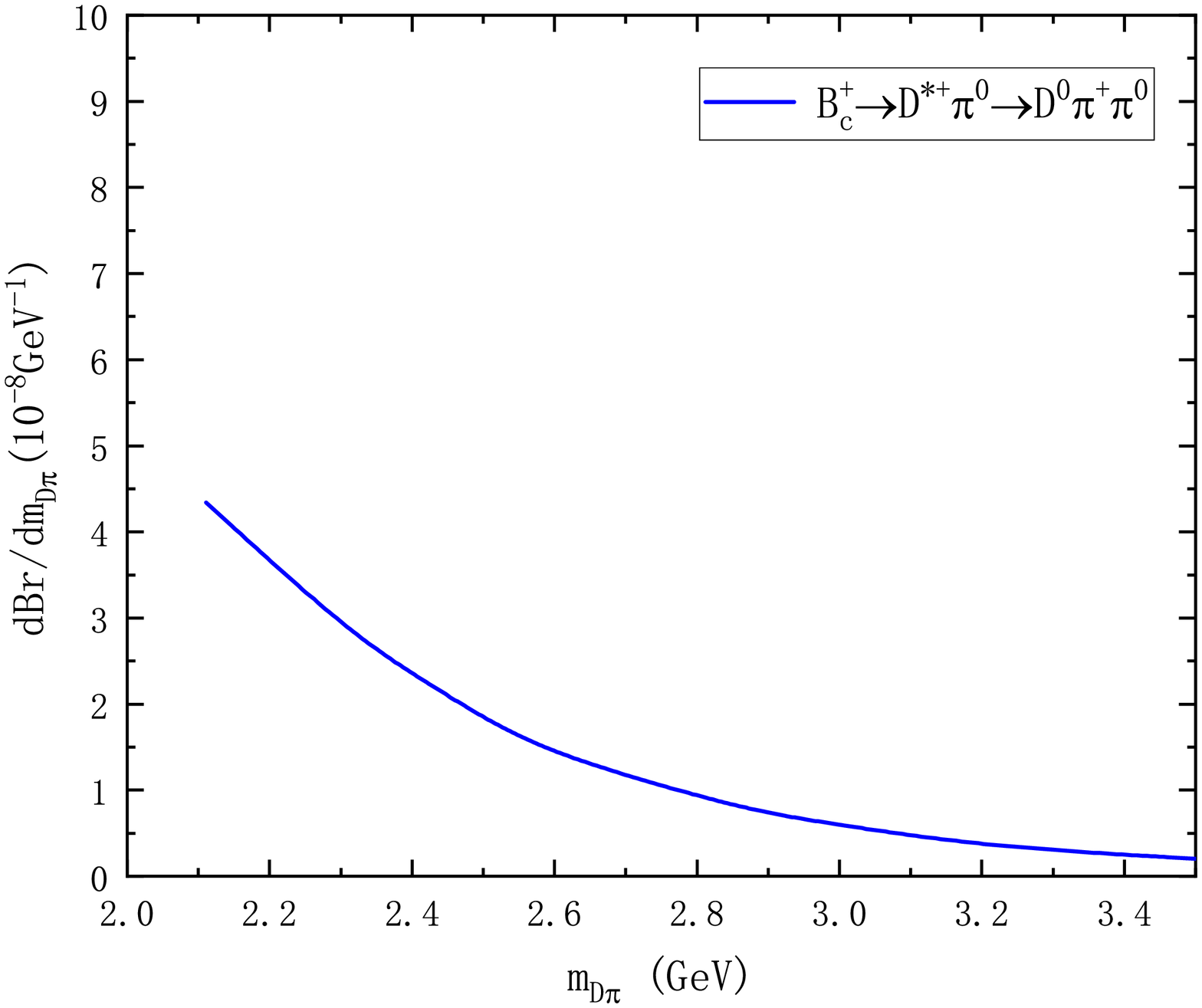}
\caption{The invariant mass $m_{D\pi}$ dependence of
the differential branching fraction for the decay $B_c\to D^{*+}\pi^{0}\to D^{0}\pi^{+} \pi^{0}$ (left panel). The corresponding
virtual contribution in the $m_{D\pi}$ region $2.1\sim3.5$ GeV (right panel).}\label{img_br1}
\end{figure}

Last, we discuss the invariant mass $m_{D\pi}$-dependence of the decay spectrum. Here we take the decay $B_{c}^+ \to D^{*+} \pi^{0}\to D^0\pi^+ \pi^{0}$
 as an example and plot the $m_{D\pi}$-dependence of
the differential branching fraction in Fig. \ref{img_br1},
where exhibits a maximum in the $D\pi$ pair invariant mass around 2.010 GeV. It is easy to see that the main contribution to the branching
ratio comes from the region around the pole mass of the $D^{*+}$ resonance as we expected. The  peak located at the $D^{*+}$ mass is very sharp,
which is caused that the decay width is tiny, $\Gamma_{D^{*+}}=83.3$ keV \cite{lee1,lee2}, at the same time, the $D^0\pi^+$ threshold is
too close to the resonance mass $m_{D^{*+}}$. It is very different with the case of the $K^{*}$ resonance \cite{pqcd3}, whose mass is much
larger than the $K\pi$ threshold.
If we integrate over $m_{D\pi}$ by limiting the range of $m_{D\pi}=[m_{D^{*+}}-\delta_m,m_{D^{*+}}+\delta_m]$ with $\delta_m=2\Gamma_{D^{*+}}, 3\Gamma_{D^{*+}}, 4\Gamma_{D^{*+}}$,
we can find that the corresponding branching factions are $93.6\%, 94.3\%, 95.0\%$ of the total branching ratio
$Br(B_{c}^+ \rightarrow D^{*+} \pi^{0}\rightarrow D^0\pi^+ \pi^{0})=0.93\times10^{-7}$. On the other hand, if we consider the
virtual contribution from the region $2.1\sim3.5$ GeV of the invariant mass $m_{D\pi}$ shown in the right panel of
Fig. \ref{img_br1}, the corresponding branching fraction is only
amount to $7.1\%$ of the total. It is simliar for the other decay channels.

{\centering\section{Summary}}

In this paper, we have studied the quasi-two-body decays $B_{c}^+\to D^{*+}h\to D^0\pi^+ h$  with $h=(K^0,\pi^0,\eta^{(\prime)})$  in the PQCD approach. The di-meson distribution
amplitude for the $D\pi$ system with the P-wave time-like form factor $F_{D\pi}(s)$ is employed to describe the $D^{*+}$ resonance and
its interactions with the $D^0\pi^+$ pair. We predict the branching
ratios of the concerned decays and find the following points:
\begin{enumerate}
\item
The branching ratios of the quasi-two-body decays $B_{c}^+\to D^{*+}h\to D^0\pi^+ h$ show an obvious hierarchy
\be
Br(B_{c}^+\to D^{*+}K^0\to D^0\pi^+ K^0)> Br(B_{c}^+\to D^{*+}\pi^0\to D^0\pi^+ \pi^0)>Br(B_{c}^+\to D^{*+}K^0\to D^0\pi^+ \eta^{(\prime)}),
\en
where $Br(B_{c}^+\to D^{*+}K^0\to D^0\pi^+ K^0)$ is the largest one and reaches up to $10^{-6}$ order. So the decay
$B_{c}^+\to D^{*+}K^0\to D^0\pi^+ K^0$ is possible to be observed by
the future LHCb experiments.
\item
Using the narrow width approximation relation and the isospin conservation $\Gamma(D^{*+}\to D^0\pi^+)/\Gamma(D^{*+}\to D\pi)=2/3$, we
can relate the branching ratios of these quasi-two-body decays $B_{c}^+\to D^{*+}h\to D^0\pi^+ h$ with those of the corresponding two-body channels $B_{c}^+\to D^{*+}h$.
Our results for the branching ratios of the decays $B_{c}^+\to D^{*+}h$ are consistent with the previous PQCD calculations within errors,
while there is considerable tension between the predictions of the PQCD approach and the RCQM for the decay
$B_{c}^+\to D^{*+}K^0$. The former is
one order of magnitude larger than the latter. It is because that the annihilation type contribution is
dominant for the decay $B_{c}^+\to D^{*+}K^0$, which is not calculable under the RCQM.
\item
The branching ratio of the decay $B_{c}^+\to D^{*+}\eta\to D^{0}\pi^+\eta$ is larger than that of
$B_{c}^+\to D^{*+}\eta^\prime\to D^{0}\pi^+\eta^\prime$, it is induced by the opposite interferences
between the amplitudes  $\mathcal{A}\left(B_{c}^+ \rightarrow D^{*+} \eta_{q}\rightarrow D^0\pi^+ \eta_{q}\right)$ and
$\mathcal{A} (B_{c}^+\to D^{*+} \eta_{s}\to$ $D^0\pi^+ \eta_{s})$.
\item
From the $D^0\pi^+$ invariant mass $m_{D^0\pi^+}$-dependences of these concerned decay spectrums, one can find that the main portions of the branching
fractions concentrate in a very small region of the $m_{D^0\pi^+}$. For example, about $94\%$ of the branching ratio of the decay
$B_{c}^+\to D^{*+}\pi^0\to D^{0}\pi^+\pi^0$ comes from the realm of $2.1$ MeV around of the $D^{*+}$ pole mass.
\end{enumerate}
\section*{Acknowledgment}

This work is partly supported by the National Natural Science
Foundation of China under Grant No. 11347030, by the Program of
Science and Technology Innovation Talents in Universities of Henan
Province 14HASTIT037, Natural Science Foundation of Henan
Province under Grant No. 232300420116.

{\centering\section{Appendix: Decay amplitudes}\label{appendix}}
\setcounter{equation}{0}
\renewcommand\theequation{A.\arabic{equation}}

In this appendix we present the PQCD factorization formulas for the amplitudes of the decays $B\to D^{*+}h\to D^0\pi^+ h$

\begin{align}
\mathcal{F}_{e}^{LL}
=&4 \sqrt{2}G_F C_F \pi f_h (\eta -1) m_{B_c}^4 \int_0^1dx_1dz\int_0^{\infty}b_1db_1b_2db_2\phi_{B_c}(x_1,b_1)\phi_{D \pi}(z,b_2,\omega)\Big\{ [\sqrt{\eta } (r_b+2 z-2)\non{}
&-2 r_b-z+1]h(\alpha_e,\beta_a,b_1,b_2)E_a(t_a)-\eta h(\alpha_e,\beta_b,b_2,b_1)E_b(t_b)\Big\},\\
\mathcal{F}_{e}^{SP}
 =&-8 \sqrt{2} G_F C_F \pi f_h r_h  m_{B_c}^4 \int_0^1dx_1dz\int_0^{\infty}b_1db_1b_2db_2\phi_B(x_1,b_1)\phi_{D \pi}(z,b_2,\omega)\Big\{[(\eta-1)rb-\sqrt{\eta}z+2 \eta(z-1) \non{}
 &+2]h(\alpha_e,\beta_a,b_1,b_2)E_a(t_a)+x_1h(\alpha_e,\beta_b,b_2,b_1)E_b(t_b)\Big\},\\
\mathcal{M}_{e}^{LL} =&16 \frac{\sqrt{3}}{3} G_F C_F \pi (\eta -1) m_{B_c}^4\int_0^1dx_1dzdx_3\int_0^{\infty}b_1db_1b_3db_3\phi_{B_c}(x_1,b_1)\phi_{D \pi}(z,b_1,\omega)\phi_h^{A}(x_3)(x_3)\Big\{ (\eta -x_1- \eta x_3 \non{}
&+x_3+ \eta z - \sqrt{\eta} z -1)h(\beta_c,\alpha_e,b_1,b_3)E_{cd}(t_c)
-(x_1+\eta  x_3-x_3+\sqrt{\eta } z-z)h(\beta_d,\alpha_e,b_1,b_3)E_{cd}(t_d)\Big\},\\
\mathcal{M}_{e}^{LR} =&-16 \frac{\sqrt{3}}{3} G_F C_F \pi r_h (\sqrt{\eta} -1)
m_{B_c}^4\int_0^1dx_1dzdx_3\int_0^{\infty}b_1db_1b_3db_3\phi_{B_c}(x_1,b_1)\phi_{D \pi}(z,b_1,\omega)\Big\{[\phi_h^{P}(x_3) (\eta +x_1\non{}
&-\eta  x_3+x_3-\sqrt{\eta } z-1)+\phi_h^{T}(x_3) \left(\eta +x_1-\eta  x_3+x_3+\sqrt{\eta } z-1\right)]h(\beta_c,\alpha_e,b_1,b_3)E_{cd}(t_c)\non{}
&-[x_1 (\phi_h^{P}(x_3)-\phi_h^{T}(x_3))+(\eta -1) x_3 (\phi_h^{P}(x_3)-\phi_h^{T}(x_3))-\sqrt{\eta } z (\phi_h^{P}(x_3)+\phi_h^{T}(x_3))]\non{}
&\times h(\beta_d,\alpha_e,b_1,b_3)E_{cd}(t_d)\Big\},\\
\mathcal{M}_{e}^{SP} =&-16 \frac{\sqrt{3}}{3} G_F C_F \pi (\eta-1)
m_{B_c}^4 \int_0^1 dx_1dzdx_3 \int_0^{\infty}b_1db_1b_3db_3\phi_{B_c}(x_1,b_1)\phi_{D \pi}(z,b_1, \omega) \Big\{[(\eta+x_1-\eta x_3+x_3 \non{}
&+\eta z-\sqrt{\eta} z-1) \phi_{h}^{A}(x_3)]h(\beta_c,\alpha_e,b_1,b_3)E_{ef}(t_e)-[(x_1+\eta x_3-x_3+\sqrt{\eta}z-z) \phi_{h}^{A}(x_3)]\non{}
&\times h(\beta_d,\alpha_e,b_1,b_3)E_{ef}(t_f)\Big\},\\
\mathcal{M}_{a}^{LL} =&16 \frac{\sqrt{3}}{3} G_F C_F \pi
m_{B_c}^4\int_0^1dx_1dzdx_3\int_0^{\infty}b_1db_1b_2db_2\phi_{B_c}(x_1,b_1)\phi_{D \pi}(z,b_2,\omega)\Big\{[(\eta -1) \phi_h^{A}(x_3) (\eta +r_b\non{}
&-\eta  x_3+x_3+\eta z -1)+\sqrt{\eta } r_h (\phi_h^{T}(x_3) (-\eta +(\eta-1)  x_3+ z+1)+\phi_h^{P}(x_3) (\eta -\eta  x_3+x_3+z-1))] \non{}
&h(\beta_e,\alpha_a,b_1,b_2)E_{ef}(t_e)-[(\eta -1)  \phi_h^{A}(x_3) (r_c+(\eta -1) (z-1))+\sqrt{\eta } r_h (\phi_h^{P}(x_3) (\eta -x_1-\eta  x_3+x_3+z-1)\non{}
&+\phi_h^{T}(x_3) (x_1+\eta  (x_3-1)-x_3+z-1))]h(\beta_f,\alpha_a,b_1,b_2)E_{ef}(t_f)\Big\},\\
\mathcal{M}_{a}^{LR} =&-16 \frac{\sqrt{3}}{3} G_F C_F \pi
m_{B_c}^4\int_0^1dx_1dzdx_3\int_0^{\infty}b_1db_1b_2db_2\phi_{B_c}(x_1,b_1)\phi_{D \pi}(z,b_2,\omega)\Big\{[(\eta -1) r_b (\sqrt{\eta } \phi_h^{A}(x_3)\non{}
&-r_h \phi_h^{P}(x_3)) +(\eta+1) r_b r_h \phi_h^{T}(x_3)+(\eta-1) r_P (x_3-1) (\phi_h^{P}(x_3)+\phi_h^{T}(x_3))+z (\eta^{3/2} \phi_h^{A}(x_3)-\sqrt{\eta} \phi_h^{P}(x_3)\non{}
&+\eta r_h (\phi_h^{T}(x_3)-\Phi_h^{P}(x_3)))]h(\beta_e,\alpha_a,b_1,b_2)E_{ef}(t_e)
-[r_h (\phi_h^{P}(x_3)+\phi_h^{T}(x_3)) (r_c+x_1-x_3)\non{}
&+\eta^{3/2} \phi_h^{A}(x_3) (r_c+z-1)-\sqrt{\eta} \phi_h^{A}(x_3) (r_c+z-1)
]h(\beta_f,\alpha_a,b_1,b_2)E_{ef}(t_f)\Big\},\\
\mathcal{F}_{a}^{LL} =&-4 \sqrt{2}G_F C_F \pi f_{B_c}  m_{B_c}^4 \int_0^1dzdx_3\int_0^{\infty}b_2db_2b_3db_3\phi_{D \pi}(z,b_2,\omega)\Big\{[(\eta -1) (z-1) \phi_h^{A}(x_3)\non{}
&-2 \sqrt{\eta } r_h z \phi_h^{P}(x_3)]h(\alpha_a,\beta_g,b_2,b_3)E_{g}(t_g)-[-(\eta -1) (\eta \phi_h^{A}(x_3)+r_c r_h \phi_h^{P}(x_3))-(\eta +1) r_c r_h \phi_h^{T}(x_3)\non{}
&+(\eta-1)^{2} x_3 \phi_h^{A}(x_3))]h(\alpha_a,\beta_h,b_3,b_2)E_{h}(t_h)\Big\},\\
\mathcal{F}_{a}^{SP} =&8 \sqrt{2}G_F C_F \pi f_{B_c} C_{D \pi} m_{B_c}^4 \int_0^1dzdx_3\int_0^{\infty}b_2db_2b_3db_3\phi_{D \pi}(z,b_2,\omega)\Big\{[2 r_h \phi_h^{P}(x_3) (\eta-z \eta -1)\non{}
&+(\eta-1) \sqrt{\eta} (z-1) \phi_h^{A}(x_3)
]h(\alpha_a,\beta_g,b_2,b_3)E_{g}(t_g)+[-(\eta -1) (\eta \phi_h^{A}(x_3)+r_c r_h \phi_h^{P}(x_3))\non{}
&-(\eta+1) r_c r_h \phi_h^{T}(x_3)+(\eta-1)^{2} x_3 \phi_h^{A}(x_3)]h(\alpha_a,\beta_h,b_3,b_2)E_{h}(t_h)\Big\},
\end{align}
where the mass ratio $r_h/m_{B_c} (r_b=m_b/m_{B_c},  r_c=m_c/m_{B_c})$ with $r_h (r_b, r_c)$ being the final bachelor meson ($b, c$ quark) mass,
$f_{B_c}$ and $f_h$ are the decay constants of $B_c$ and the final bachelor meson $h=(K, \pi, \eta^{(\prime)})$, respectively.
The hard scales are chosen as

\be{}
\bea{}
t_a &= max\{\sqrt{|\beta_a|},1/b_1,1/b_2\},\;\;\;\;\;\;\;\;\;\;\;\;\;\;
t_b = max\{\sqrt{|\beta_b|},1/b_1,1/b_2\},\\
t_c &= max\{\sqrt{|\beta_c|},\sqrt{|\alpha_e|},1/b_1,1/b_3\},\;\;\;
t_d = max\{\sqrt{|\beta_d|},\sqrt{|\alpha_e|},1/b_1,1/b_3\},\\
t_e &= max\{\sqrt{|\beta_e|},\sqrt{|\alpha_a|},1/b_1,1/b_2\},\;\;\;
t_f = max\{\sqrt{|\beta_f|},\sqrt{|\alpha_a|},1/b_1,1/b_2\},\\
t_g &= max\{\sqrt{|\beta_g|},\sqrt{|\alpha_a|},1/b_2,1/b_3\},\;\;\;
t_h = max\{\sqrt{|\beta_h|},\sqrt{|\alpha_a|},1/b_2,1/b_3\},
\ena{}
\en{}
where
\be{}
\bea{}
\alpha_e &= x_1 z m_{B_c}^2, \;\;\;\alpha_a = (z-1)[x_3+(1-x_3) \eta] m_{B_c}^2,\\
\beta_a &= (z+r_b^2-1)m_{B_c}^2,\;\;\; \beta_b = x_1m_B^2,\\
\beta_c &= [x_1 z -z (1-x_3) (1-\eta)]m_{B_c}^2,\;\;\;\beta_d = [x_1 z -z x_3 (1-\eta)]m_{B_c}^2,\\
\beta_e &= [r_b^2-z(1-x_3)(1-\eta)]m_{B_c}^2,\;\;\; \beta_f = [r_c^2-(z-1)(x_1-x_3(1-\eta)-\eta)]m_{B_c}^2,\\
\beta_g &= (z-1)m_{B_c}^2,\;\;\; \beta_h = [r_c^2-(\eta+x_3(1-\eta))]m_{B_c}^2.
\ena{}
\en{}

The hard functions are written as
\be{}
\bea{}
h(\alpha, \beta, b_1, b_2) &= h_1(\alpha,b_1)\times h_2(\beta,b_1,b_2), \\
h_1(\alpha,b_1) &=\Big\{\begin{array}{cc}
                    K_0(\sqrt{\alpha}b_1), & \alpha > 0 \\
                    K_0(i \sqrt{-\alpha}b_1), & \alpha < 0
                  \end{array}\\
h_{2}\left(\beta, b_{1}, b_{2}\right)&=\left\{\begin{array}{ll}
\theta\left(b_{1}-b_{2}\right) I_{0}\left(\sqrt{\beta} b_{2}\right) K_{0}\left(\sqrt{\beta} b_{1}\right)+\left(b_{1} \leftrightarrow b_{2}\right), & \beta>0 \\
\theta\left(b_{1}-b_{2}\right) J_{0}\left(\sqrt{-\beta} b_{2}\right) K_{0}\left(i \sqrt{-\beta} b_{1}\right)+\left(b_{1} \leftrightarrow b_{2}\right), & \beta<0
\end{array}\right.
\ena{}
\en{}
with
\be{}
\bea{}
K_0(ix) = \frac{\pi}{2}(-N_0(x)+iJ_0(x)).
\ena{}
\en{}

The Sudakov factor $S_t(x)$ from the threshold resummation is given as
\be{}
S_t(x) = \frac{2^{1+2a}\Gamma(3/2+a)}{\sqrt{\pi}\Gamma(1+a)}[x(1-x)]^a,
\en{}
with the parameter $a = 0.4$. The evolution factors are given by
\be
E_{a}(t) &=& \alpha_s(t)exp[-S_{B_c}(t)-S_{D^*}(t)]S_t(z), \;\;\;E_{b}(t) = \alpha_s(t)exp[-S_{B_c}(t)-S_{D^*}(t)]S_t(x_1),\\
E_{g}(t) &=& \alpha_s(t)exp[-S_{D^*}(t)-S_h(t)]S_t(x_3), \;\;\;E_{h}(t) = \alpha_s(t)exp[-S_{D^*}(t)-S_h(t)]S_t(z),\\
E_{cd}(t) &=& \alpha_s(t)exp[-S_{B_c}(t)-S_{D^*}(t)-S_h(t)]|_{b_2=b_1},\\
E_{ef}(t) &=& \alpha_s(t)exp[-S_{B_c}(t)-S_{D^*}(t)-S_h(t)]|_{b_3=b_2},
\en
where the Sudakov exponents are listed as
\be{}
\bea{}
S_{B_c}(t) =& s(x_1\frac{m_{B_c}}{\sqrt{2}},b_1)+\frac{5}{3}\int_{1/b_1}^t\frac{d\Bar{\mu}}{\Bar{\mu}}\gamma_q(\alpha_s(\Bar{\mu})),\\
S_{D^*}(t) =& s(z\frac{m_{B_c}}{\sqrt{2}},b_2)+s((1-z)\frac{m_{B_c}}{\sqrt{2}},b_2)+2\int_{1/b_2}^t\frac{d\Bar{\mu}}{\Bar{\mu}}\gamma_q(\alpha_s(\Bar{\mu})),\\
S_{h}(t)=& s\left(x_{3} \frac{m_{B_c}}{\sqrt{2}}, b_{3}\right)+s\left(\left(1-x_{3}\right) \frac{m_{B_c}}{\sqrt{2}}, b_{3}\right)+2 \int_{1 / b_{3}}^{t} \frac{d \bar{\mu}}{\bar{\mu}} \gamma_{q}\left(\alpha_{s}(\bar{\mu})\right),
\ena{}
\en{}
with $\gamma_q = -\alpha_s/\pi$ being the quark anomalous dimension. The function $s(Q,b)$ is expressed as
\be{}
\bea{}
s(Q,b) =& \frac{A^{(1)}}{2\beta_1}\hat{q}ln\left(\frac{\hat{q}}{\hat{b}}\right)-\frac{A^{(1)}}{2\beta_1}(\hat{q}-\hat{b})+\frac{A^{(2)}}{4\beta_1^2}\hat{q}ln\left(\frac{\hat{q}}{\hat{b}}-1\right)\\
&-[\frac{A^{(2)}}{4\beta_1^2}-\frac{A^{(1)}}{4\beta_1}ln(\frac{e^{2\gamma_E-1}}{2})]ln(\frac{\hat{q}}{\hat{b}}),
\ena{}
\en{}
where
\be{}
\bea{}
\hat{q} =& ln\frac{Q}{\sqrt{2}\Lambda_{QCD}},\hat{b} = ln\frac{1}{b\Lambda_{QCD}},\\
\beta_1 =& \frac{33-2n_f}{12},\beta_2 = \frac{153-19n_f}{24},\\
A^{(1)} =& \frac{4}{3}, A^{(2)} = \frac{67}{9}-\frac{\pi^2}{3}-\frac{10}{27}n_f+\frac{8}{3}\beta_1ln(\frac{1}{2}e^{\gamma_E}),
\ena{}
\en{}
with $n_f$ being the number of the quark flavor, and $\gamma_E$ the Eular constant.


\begin{thebibliography}{99}
\bibitem{shiyan1}
R. Aajj \textit{et al.} [LHCb Collaboration], Phys. Rev. Lett. {\bf108},
251802 (2012) [arXiv:1204.0079 [hep-ex]].
\bibitem{shiyan2}
 R. Aaij \textit{et al.} [LHCb Collaboration],  Phys.
Rev. Lett. {\bf109}, 232001 (2012) [arXiv:1209.5634 [hep-ex]].
\bibitem{shiyan3}
R. Aajj \textit{et al.} [LHCb Collaboration], JHEP {\bf09}, 075 (2013) [arXiv:1306.6723 [hep-ex]].
\bibitem{shiyan4}
R. Aajj \textit{et al.} [LHCb Collaboration], Phys. Rev. D {\bf87},
112012 (2013) [arXiv:1304.4530 [hep-ex]].
\bibitem{shiyan5}
R. Aajj \textit{et al.} [LHCb Collaboration], JHEP {\bf11}, 094 (2013) [arXiv:1309.0587 [hep-ex]].
\bibitem{shiyan6}
R. Aajj \textit{et al.} [LHCb Collaboration], Phys. Rev. Lett. {\bf111},
181801 (2013) [arXiv:1308.4544 [hep-ex]].
\bibitem{shiyan7}
R. Aajj \textit{et al.} [LHCb Collaboration], JHEP {\bf09}, 153 (2016) [arXiv:1607.06823 [hep-ex]].
\bibitem{shiyan8}
G. Aad \textit{et al.} [ATLAS Collaboration],  JHEP {\bf08}, 087 (2022) [arXiv:2203.01808 [hep-ex]].
\bibitem{sym1}
B. Bhattacharya, M. Gronau, J.L. Rosner, Phys. Lett. B {\bf726}, 337 (2013)  [arXiv:1306.2625 [hep-ph]].
\bibitem{sym4} 
M. Gronau, Phys. Lett. B {\bf727}, 136 (2013) [arXiv:1308.3448 [hep-ph]].
\bibitem{sym6}
D. Xu, G. N. Li, and X. G. He, Phys. Lett. B {\bf728}, 579 (2014) [arXiv:1311.3714 [hep-ph]].
\bibitem{sym5}
M. Gronau, J.L. Rosner, Phys. Rev. D {\bf72}, 094031 (2005) [arXiv:hep-ph/0509155].
\bibitem{sym2} 
G. Engelhard, G. Raz, Phys.Rev. D {\bf72}, 114017 (2005) [arXiv:hep-ph/0508046].
\bibitem{sym3} 
M. Imbeault, D. London, Phys. Rev. D {\bf84}, 056002 (2011) [arXiv:1106.2511 [hep-ph]].
\bibitem{fat3}
S. H. Zhou, R. H. Li, Z. Y. Wei and C. D. Lu, \prd {\bf104}, 116012 (2021) [arXiv:2107.11079 [hep-ph]].
\bibitem{qcdf1}
Z. H. Zhang, X. H. Guo, Y. D. Yang, Phys. Rev. D {\bf87}, 076007 (2013) [arXiv:1303.3676 [hep-ph]].
\bibitem{qcdf2}
H. Y. Cheng, C. K. Chua, Z. Q. Zhang, Phys. Rev. D {\bf94}, 094015 (2016) [arXiv:1607.08313 [hep-ph]].
\bibitem{qcdf5}
S. K$\ddot{r}$ankl, T. Mannel, J. Virto, Nucl. Phys. B {\bf899}, 247 (2015) [arXiv:1505.04111 [hep-ph]].
\bibitem{qcdf6}
R. Klein, T. Mannel, J. Virtob, K. Keri Vos, JHEP {\bf10}, 117 (2017) [arXiv:1708.02047 [hep-ph]].
\bibitem{pqcd1}
W. F. Wang and H. n. Li, Phys. Lett. B {\bf763}, 29 (2016) [arXiv:1609.04614 [hep-ph]].
\bibitem{pqcd2}
Z. Q. Zhang, H. x. Guo, Eur.Phys.J.C {\bf79}, 59 (2019) [arXiv:1812.11372 [hep-ph]].
\bibitem{pqcd3}
Z. Q. Zhang, Y. C. Zhao, Z. L. Guan, Z. J. Sun and Z. Y. Zhang, Chin. Phys. C {\bf46}, 123105 (2022) [arXiv:2207.02043 [hep-ph]].
\bibitem{pqcd4}
C. Wang, J. B. Liu, H. n. Li, C. D. Lu, Phys. Rev. D {\bf97}, 034033 (2018) [arXiv:1711.10936 [hep-ph]].
\bibitem{pqcd5}
A. J. Ma, Y. Li, Z. J. Xiao, Nucl.Phys. B {\bf926} 584 (2018) [arXiv:1710.00327 [hep-ph]].
\bibitem{pqcd6}
Z. Rui, Y. Li, H. n. Li, Phys. Rev. D {\bf98}, 113003 (2018) [arXiv:1809.04754 [hep-ph]].
\bibitem{pqcd7}
Y. Li, W. F. Wang, A. J. Ma, Z. J. Xiao, Eur. Phys. J. C {\bf79}, 37 (2019) [arXiv:1809.09816 [hep-ph]].
\bibitem{pqcd8}
W. F. Wang, Phys. Let. B {\bf788} 468 (2019) [arXiv:1809.02943 [hep-ph]].
\bibitem{pqcd9}
Z. T. Zou, W. S. Fang, X. Liu, Y. Li, Eur. Phys. J. C {\bf82}, 1076 (2022) [arXiv:2210.08522 [hep-ph]].
\bibitem{pqcd10}
Y. Li, D. C. Yan, J. Hua, Z. Rui, H. n. Li, Phys. Rev. D {\bf104}, 096014 (2021) [arXiv:2105.03899 [hep-ph]].
\bibitem{pqcd11}
Z. Rui, Y. Li, W. F. Wang, Eur. Phys. J. C {\bf77}, 199 (2017) [arXiv:1701.02941 [hep-ph]].
\bibitem{canshu1}
Z. Rui, Z. T. Zou, C.D. Lu, Phys. Rev. D {\bf86}, 074008 (2011) [arXiv:1112.1257 [hep-ph]].
\bibitem{rcqm1}
J.F. Liu, K.T. Chao, Phys. Rev. D {\bf56}, 4133 (1997).
\bibitem{cff1}
C. H. Chen and H. n. Li, Phys. Lett. B {\bf561}, 258 (2003) [arXiv:hep-ph/0209043].
\bibitem{cff2}
C. H. Chen and H. n. Li, Phys. Rev. D {\bf70}, 054006 (2004) [arXiv:hep-ph/0404097].
\bibitem{DstADs1}
W .F .Wang, J. Chai, Phys. Lett. B {\bf791}, 342-350 (2019) [arXiv:1812.08524 [hep-ph]].
\bibitem{kw}
K. M. Watson, Phys. Rev. {\bf88}, 1163 (1952).
\bibitem{rbw}
G. Breit, E. Wigner, Phys. Rev. {\bf49}, 519 (1936).
\bibitem{blatt}
 J. Blatt, V. Weisskopf, Theoretical Nuclear Physics, (John Wiley $\&$ Sons, New York, 1952).
\bibitem{rbw2}
R. Aaij \textit{et al.} [LHCb Collaboration], Phys. Rev. D {\bf94}, 072001 (2016) [arXiv:1608.01289 [hep-ex]].
\bibitem{rbw3}
R. Aaij \textit{et al.} [LHCb Collaboration], Phys. Rev. D {\bf91}, 092002 (2015) [arXiv:1503.02995 [hep-ex]].
\bibitem{PADs1}
H. n. Li, S. Mishima, A. I. Sanda, Phys. Rev. D {\bf72}, 114005 (2005) [arXiv:hep-ph/0508041].
\bibitem{PADs2}
P. Ball and R. Zwicky, Phys. Rev. D {\bf71}, 014015 (2005) [arXiv:hep-ph/0406232].
\bibitem{PADs3}
Z. J. Xiao, Z. Q. Zhang, X. Liu and L. B. Guo, Phys. Rev. D {\bf78}, 114001 (2008) [arXiv:0807.4265 [hep-ph]].
\bibitem{hua}
J. Hua, M. H. Chu, J. C. He, X. Ji, A. Sch$\ddot{a}$fer, Y. Su, P. Sun, W. Wang, J. Xu, Y. B. Yang, F. Yao, J. H. Zhang, Q. A. Zhang,
Phys. Rev. Lett. {\bf129}, 132001 (2022).
\bibitem{pball}
P. Ball, V. M. Braun and A. Lenz, JHEP {\bf0605}, 004 (2006) [arXiv:hep-ph/0603063].
\bibitem{xliu}
X. Liu, H. n. Li and Z. J. Xiao, Phys. Rev. D {\bf97}, 113001 (2018) [arXiv:1811.12738 [hep-ph]].
\bibitem{heff2}
G. Buchalla, A. J. Buras, M. E. Lautenbacher, Rev. Mod. Phys. {\bf68}, 1125 (1996) [arXiv:hep-ph/9512380].
\bibitem{pdg}
R. L. Workman \textit{et al.} [Particle Data Group], Review of Particle
Physics, PTEP {\bf2022}, 083C01 (2022).
\bibitem{feld}
T. Feldmann, P. Kroll, and B. Stech, Phys. Rev. D {\bf58}, 114006 (1998) [arXiv:hep-ph/9802409].
\bibitem{br1}
B. El-Bennich, A. Furman, R. Kaminski et al. Phys. Rev. D {\bf74}, 114009 (2006)[arXiv:hep-ph/0608205].
\bibitem{br2}
B. El-Bennich, A. Furman, R. Kaminski et al. Phys. Rev. D {\bf79}, 094005 (2009) [arXiv:0902.3645 [hep-ph]].
\bibitem{lhcb}
R. Aaij \textit{et al.} [LHCb Collabration], Phys. Rev. Lett. {\bf118}, 111803 (2017) [arXiv:1701.01856[hep-ex]].
\bibitem{lhcb3}
R. Aaij \textit{et al.} [LHCb Collabration], Phys. Rev. Lett. {\bf109}, 232001 (2012) [arXiv:1209.5634[hep-ex]].
\bibitem{lee1}
J.P. Lees \textit{et al.} [BaBar Collaboration], Phys. Rev. D {\bf88}, 052003 (2013) [\textit{Erratum ibid.} Phys. Rev. D {\bf88},  079902 (2013)] [arXiv:1304.5009 [hep-ex]].
\bibitem{lee2}
J.P. Lees \textit{et al.} [BaBar Collaboration], Phys. Rev. Lett. {\bf111}, 111801 (2013) [\textit{Erratum ibid.} Phys. Rev. Lett. {\bf111}, 169902 (2013)] [arXiv:1304.5657 [hep-ex]].
\end{thebibliography}
\end{document}